\documentstyle[12pt]{article}
\input epsf
\begin{document}
\title{
\hfill{\small TPBU-95-6}
\vspace*{0.5cm}\\
\sc
Critical exponent of the localization length for the 
 symplectic case
\vspace*{0.3cm}}
\author{\sc Alexander Moroz\thanks{e-mail address :
{\tt am@th.ph.bham.ac.uk}}
\thanks{On leave from Institute of Physics, Na Slovance 2,
CZ-180 40 Praha 8, Czech Republic}
\vspace*{0.3cm}}
\date{
\protect\normalsize
\it School of Physics and Space Research, 
University of Birmingham,
Edgbaston, Birmingham B15 2TT, U. K.
}
\maketitle
\begin{center}
{\large\sc abstract}
\end{center}
A new summability method was tested to calculate the critical exponent
$\nu$ of the localization length for the symplectic case
derived from the non-linear $\sigma$-model. 
Although we used the same series as  Hikami and others, unlike them
we were able to resum the series in two-dimensions (2D)
and obtain the  result $\nu\sim 1$. Values of $\nu$ in $2+\varepsilon$ 
dimensions seem to saturate the Harris inequality
up to $\varepsilon=0.2$. 
\vspace*{0.2cm}


\vspace{3.2cm}

\begin{center}
{\bf (J. Phys. A: Math. Gen. 29, 289-294 (1996))}
\end{center}
\thispagestyle{empty}
\baselineskip 20pt
\newpage
\setcounter{page}{1}
{\bf 1.} {\em Introduction}.-
%
Anderson localization is known as a problem where the wave function
localizes due to a random potential scattering \cite{A}.
The critical behaviour of the localization transition is described by the
non-linear (NL) $\sigma$-model \cite{Weg}.
The NL$\sigma$-model explains succesfully various aspects of the
Anderson localization, including the nonsingular density of states and three different universality classes - the orthogonal, unitary, and symplectic,
depending on whether time-reversal symmetry is preserved
or spin-flip scattering occurs. 

The critical exponent $\nu$ describes the behaviour of the 
localization length
$\lambda$ near the mobility edge $E_c$,
\begin{equation}
\lambda \sim (E_c-E)^{-\nu}.
\end{equation}
To obtain critical exponents at the transition point in the
framework of the NL$\sigma$-model,
 the scheme of the minimal subtraction by the dimensional regularization
is usually used. The critical
exponent $\nu$ of the correlation length or the localization length
is then given by the relation 
\begin{equation}
\nu=-\frac{1}{\beta'(t_c)},
\label{btnurel}
\end{equation}
where $t_c$ is the zero of the $\beta$ function and $\beta'$ is t
he derivative of
$\beta$ with respect to $t$ \cite{BZ}. In general, the $\beta$ function 
of the  NL$\sigma$-model in $2+\varepsilon$ dimensions is given as
\begin{equation}
\beta(t)=\varepsilon\, t -\sum_{n=2}^\infty a_n t^{n},
\label{btform}
\end{equation}
where the coefficients $a_n{}'s$ do not depend on $\varepsilon$ and
$a_2\neq 0$ \cite{Hi2}.
In the symplectic case, which is the universality class
corresponding to time-reversal
symmetry and  strong spin-orbit coupling, the $\beta$ function to 
five-loop order is given by
\begin{equation}
\beta_{so}(t)=\varepsilon\, t +t^2 -\frac{3}{4}\zeta(3)\,t^5 -
\frac{27}{64}\zeta(4)\,t^6 + {\cal O} (t^7),
\label{bts}
\end{equation}
where $\zeta$ is the Riemann $\zeta$-function \cite{Hi2}.

{\bf 2.} {\em Methods}.-
To extract the critical exponent $\nu$, only the Borel-Pad\'{e}
method has been used so far \cite{Hi2}.
The Borel summability method, when applied to the series
of the form (\ref{btform}), consists in the following
transformation,
\begin{equation}
\sum_{n=1}^\infty a_n t^{n} 
\longrightarrow \frac{1}{t}
\int_0^\infty e^{-u/t} \sum_{n=1}^\infty a_n \frac{u^{n}}{n!}\,du.
\label{botr}
\end{equation}
For finite series, relation (\ref{botr})
is useless since it is an identity. One can integrate term by term, by using
\[
\frac{1}{t}\int_0^\infty e^{-u/t}u^n du =n!\, t^n.
\]
However, for infinite
series the second form in (\ref{botr})
may have a much wider range of applicability.
For example, in the case of the series $\sum_{n=0}^\infty z^n$
the second form  converges in the whole complex 
halfplane Re $z<1$ despite that the original series being divergent
for $|z|>1$, outside its radius of convergence.
Having   only a finite number of a series up to order 
$N$ at one's disposal, such as in the present case, 
the integrand in (\ref{botr}) can be approximated by
the Pad\'{e} approximation which generates 
an infinite power series coinciding up to order $N$
with the original series. The resulting method is called
the Borel-Pad\'{e} method.

\begin{figure}
\centerline{\epsfxsize=8cm \epsfbox{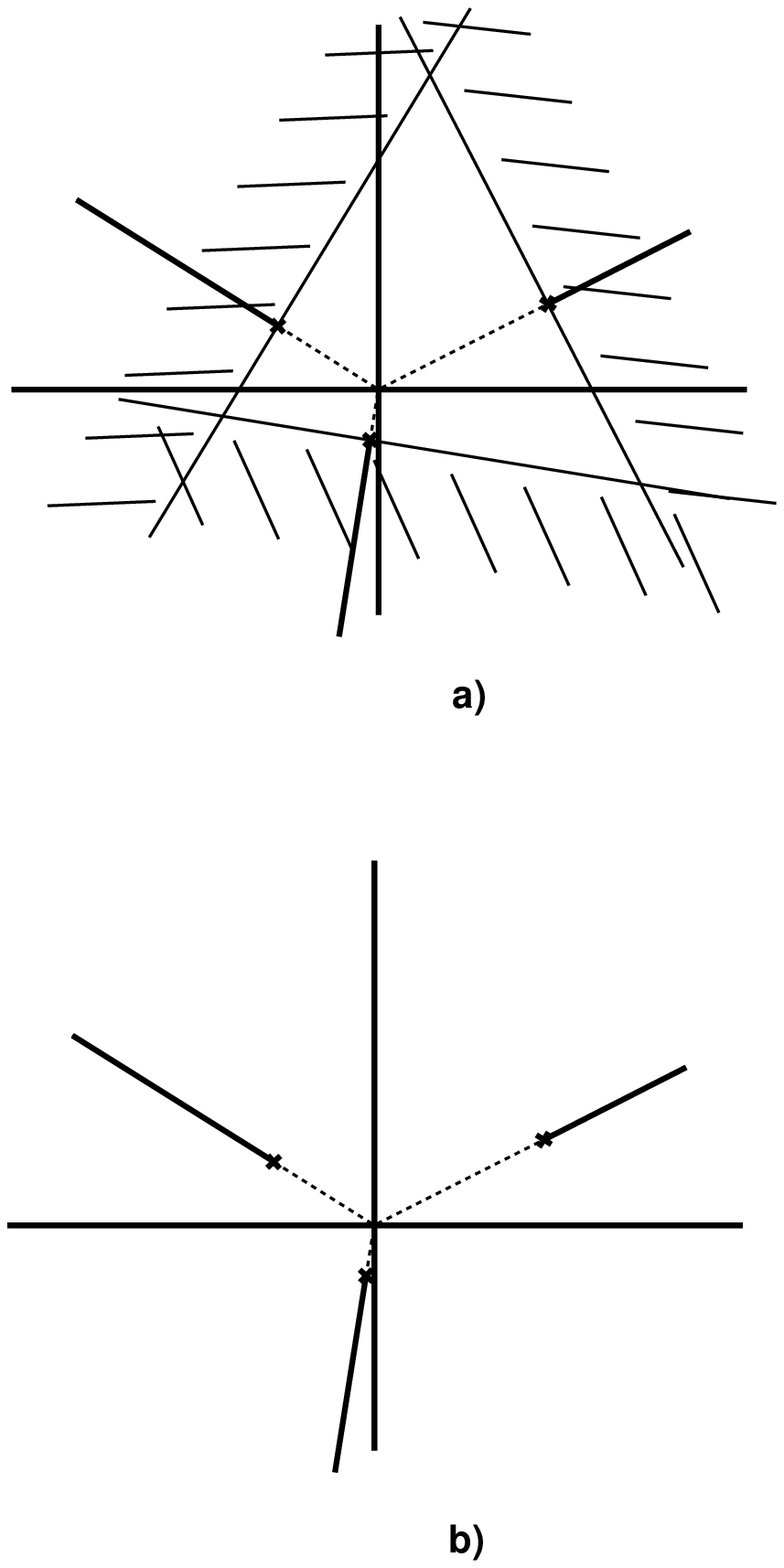}}
\caption{{\bf a)} The actual region of convergence 
for the Borel method in the case of the Taylor series of an 
analytic function $f(z)$
is a polygon which is obtained by removing
half-planes from the complex plane  which lie behind
a perpendicular to the ray from the origin 
passing through the singularity (a cross).
{\bf b)} In the case of our method, the actual region of convergence is 
obtained by  only removing the part of the 
ray behind each singularity. Such region is called
the Mittage-Leffler star.
}
\label{mls}
\end{figure}
Several years ago we have developed a method \cite{Mor} which
consists in a similar transformation as (\ref{botr}) 
but with $e^{-u/t}$ replaced by $e^{-e^{u/t}}$ and  $n!$ replaced
by
\begin{equation}
\mu(n) =\int_0^\infty e^{-e^u} u^n\, du. 
\end{equation}
Moments $\mu(n)$ are increasing more slowly then $n!$
(they behave roughly as $\ln ^n(n)$ when $n\rightarrow\infty$), but
 as far as analytic properties are concerned 
this results in a wider region of convergence.
If $f(z)$ denotes an analytic continuation of a power series 
with a non-zero radius of convergence, then 
our method gives a finite result $f(z)$ in the
so-called Mittag-Leffler star of $f(z)$ \cite{Mor}. 
In the case of the series
$\sum_{n=0}^\infty z^n$ it means that our method converges in the 
whole complex plane except for the interval $[1,\infty)$.
For a general power series, the Mittag-Leffler star is obtained
in the following way (see Fig. \ref{mls}).
First, one draws rays  from
the origin and passing through singularities of $f(z)$. 
The Mittage-Leffler
star is the region which remains in the complex plane after
the part of the ray beyond each singularity is removed. 
We recall that  the actual region of convergence 
for the Borel method is a polygon which is obtained by removing
half-planes from the complex plane  which lie behind
a perpendicular to the ray from the origin
passing through the singularity
(see Fig. \ref{mls}). 
If one has only a finite number of terms of a series at disposal, 
one can use the Pad\'{e} approximation of the integrand in (\ref{botr})
in the same way as when the Borel method is used.
We shall call the resulting method the $\mu$-Pad\'{e} method.
Both, the Borel method and our method are so-called analytic
 moment-constant summability methods \cite{Mor,H}. 
They can be applied to both convergent
and divergent series. 

\begin{figure}
\centerline{\epsfxsize=8cm \epsfbox{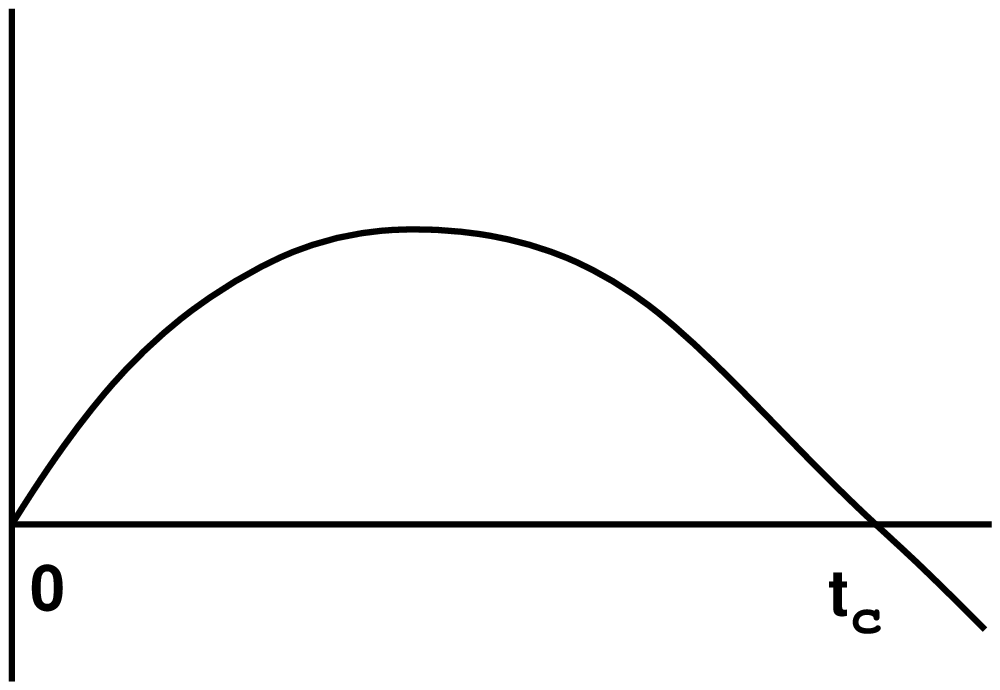}}
\caption{Typical behaviour of the $\beta$ function for the
symplectic case after it has been resummed by the
diagonal $\mu$-Pad\'{e} method.
}
\label{beta}
\end{figure}
{\bf 3.} {\em Results}.-
One of the motivations in using  the $\mu$-Pad\'{e} method
for the symplectic case was the fact 
the series (\ref{bts}) is not
Borel summable and the Borel-Pad\'{e} method does not work
for this case \cite{Hi2}.
In the $2D$ symplectic case, to leading order, so-called
weak anti-localization occurs \cite{OJ} and until recently it appeared
that this result remain unchanged by higher order terms.
This included the strange result that there is  no fixed point
(and hence no transition and no localization) for
the $2D$ symplectic case \cite{KM}.
We have applied the  diagonal [3/3] $\mu$-Pad\'{e} method directly to the
$\beta_{so}(t)$  and looked for its zero $t_c(\varepsilon)$
as a function of $\varepsilon$ (see Fig.\ \ref{beta}). 
By using relation
(\ref{btnurel}) we have found that as 
$\varepsilon\downarrow 0$, the critical exponents
\begin{equation}
\nu\rightarrow 0.98 
\label{2dnu}
\end{equation}
in the $2D$ symplectic  case.
The $\beta$ function for the orthogonal universality class
is related to $\beta$ function for the symplectic 
universality class by substituting $t$ in Eq.\ (\ref{bts})
by $-2t$ \cite{Hi2}.
Therefore, because the diagonal [3/3] $\mu$-Pad\'{e} method works
for the symplectic case, it cannot be applied to the orthogonal case,
since the Pad\'{e} approximant, having a polynomial of order
$3$ in the denominator, develops a pole in the integration interval.
A similar statement applies to
the diagonal [3/3] Borel-Pad\'{e} method, which, in contrast, works
for the orthogonal case
and therefore does not for the symplectic case \cite{Hi2}.

In the following, we have scanned the $\varepsilon$-expansion by varying
$\varepsilon=d-2$, where $d$ is the space dimension, within the interval
$\varepsilon\in(0,1]$. Results are presented in Table I.
\begin{center}
TABLE I. $\beta'(t_c)$ and the critical coefficient $\nu$
as a function of $\varepsilon$
\vspace*{0.3cm}\\
\begin{tabular}{||c||c|c|c|c|c|c|c|c|c|c||} \hline\hline
& & & & &&&&&& \\ 
$\varepsilon$ & 0.1 & 0.2 &0.3&0.4&0.5&0.6&0.7&0.8&0.9&1.0\\
& & & & &&&&&& \\  \hline\hline
& & & & &&&&&& \\ 
$\beta'(t_c)$ &-1.05 &-1.14 & -1.27&-1.47&
-1.74& -2.1&-2.58&-3.2& & -0.50  \\ 
& & & & &&&&&& \\ \hline 
& & & & &&&&&& \\ 
$\nu$ &0.95 &0.88 & 0.79& 0.68& 0.57& 0.48 &
0.39& 0.31 & & 2 \\
& & & & &&&&&& \\ \hline\hline
\end{tabular}
\vspace*{0.3cm}
\end{center}

{\bf 4.} {\em Discussion}.-
We have checked whether our results satisfy the Harris inequality 
\cite {Har},
\begin{equation}
\nu \geq \frac{2}{d},
\label{chay}
\end{equation}
derived under the assumption of the validity of  one-parameter scaling.
Up to an error due to the finite number of terms for
$\beta_{so}(t)$  [see (\ref{bts})], our result 
seems to saturate the inequality (\ref{chay}) up to $\varepsilon=0.2$.
As $\varepsilon$ increases further, $\nu$ ceases to 
satisfy (\ref{chay})
and eventually around $\varepsilon=0.9$ 
the  diagonal $\mu$-Pad\'{e} method collapses. The reason is that the
Pad\'{e} approximant develops a pole on the integration interval.
For $\varepsilon=1$, the $\mu$-Pad\'{e} method starts to work again 
and our result
\begin{equation}
\nu=2
\label{3dnu}
\end{equation}
satisfies inequality (\ref{chay}).
However, because of the collapse of the  diagonal $\mu$-Pad\'{e} method
at $\varepsilon=0.9$, this result must be taken with some reservations.
Since the $\epsilon$ expansion (\ref{btform}) is an asymptotic expansion,
it is difficult to make an extrapolation too far away from
the limiting point ($D=2$ in our case), given that the
extrapolation is based on incomplete knowledge
of the $\beta$ function.
Nevertheless, it provides a substational improvement
over previous analytical results, and there is always a chance that 
further terms of the $\beta$ function will make our results better.
It is worthwhile to mention that the actual convergence or divergence
of an asymptotic series is not as important for applicability
of an analytic summability method as whether asymptotic series
obey the {\em strong asymptotic conditions} (SAC) 
\cite{Mor,Mor1}. The latter
ensure that there exists only one function with the required
analytic properties and a given asymptotic
expansion. For example, given a convergent asymptotic
series $\sum_0^\infty z^n$ in the right complex half plane,
without the validity of the SAC the sum of this series can be
any function of the form $1/(1-z) +As^B e^{-C/z}$ with
$A$, $B$, and $C>0$ arbitrary constants. 

In the $2D$ symplectic case, our result (\ref{2dnu}) 
for the critical exponent $\nu$ is smaller
than $\nu\sim 2.5$ \cite{Ev1} or  $\nu\sim 2.74$ \cite{Fas}
obtained by numerical scaling
analysis using, respectively, the Evangelou-Ziman model \cite{EZ} 
or Ando's model \cite{An}. A similar disagreement between field 
theory predictions and tight-binding scaling methods is also
known to exist for the $3D$ orthogonal case, where the former
yields $\nu\sim 1$ and the latter $\nu\sim 1.4$.
It is interesting to note that a disagreement also exists
for the results for the critical exponent $\nu$ 
obtained by the numerical scaling analysis and that
obtained by the best fit to the critical level distribution $P(s)$ 
in the $2D$ symplectic case  at the mobility edge \cite{AKL},
\begin{equation}
P(s)=Bs^4 \exp\left(-As^\gamma\right).
\label{crit-d}
\end{equation}
Here parameter $\gamma$  is given by 
\begin{equation}
\gamma= 1- \frac{1}{\nu^* d},
\label{gm}
\end{equation}
$d$ is the dimensionality of the system, $A$ is a numerical factor 
which depends on the dimensionality, and $B$ is to be 
found from the normalization conditions \cite{AKL}. 
The critical exponent $\nu^*$ in (\ref{gm}) should be identical to
$\nu$. However recent numerical analysis implies
that $\nu^* = 0.7\pm 0.08$ \cite{SZ} or $\nu^*\sim 0.77$ \cite{OO}
for Ando's model \cite{An}, and $\nu^* = 0.83\pm 0.7$ \cite{Ev1}
for the Evangelou-Ziman model \cite{EZ}.
A comparison with (\ref{chay}) shows that $\nu^*$ even violates
the Harris inequality \cite{Har} [however $\nu^*$ satisfies a weaker 
relation, $\nu^*>1/d$, which suffices  to derive (\ref{crit-d})].
Surprisingly enough, our result (\ref{2dnu}) for the critical exponent
$\nu$ in the $2D$ symplectic case for the localization length
is actually very close to $\nu^*$ and it seems to be tempting to say
that the critical exponent obtained from the NL$\sigma$ is just
$\nu^*$. That, however, would be premature, because in the
$3D$ orthogonal case the overall behaviour of $P(s)$ has been claimed
to be well fitted with $\nu$ obtained from the numerical scaling
analysis \cite{Ev2}.

To conclude, we have demonstrated that the $\mu$-Pad\'{e} method 
can be useful in determining critical exponents. In principle,
our method can be used as an alternative to the Borel-Pad\'{e}
summability method, whenever a result is obtained in a form
of a power (asymptotic) series, such as in the case of high
temperature expansion, weak coupling expansion, etc. 
In connection with disordered systems, it would be interesting
to use our method for a calculation of the density of states and 
the diffusion constant for the problem of an electron  moving
in two dimensions in the lowest Landau level and 
in a random potential, which has been
analyzed by the Borel-Pad\'{e} method \cite{SC}.
It was suggested 
that our summability method be used
for the location of critical points \cite{Mor1},
since it finds singularities of an analytic function much more 
precisely than the Borel method (cf. Fig. \ref{mls}). 
An application of the $\mu$-Pad\'{e} method to other problems
will be given elsewhere.

I should like to thank I. V. Lerner for  discussion and suggestions on the
literature and M. W. Long for his help with computer facilities.
I should also like to thank referees for useful suggestions.
This work was supported by EPSRC grant number GR/J35214.
Partial support by the Grant Agency of the Czech Republic under  
Project No. 202/93/0689 is also gratefully acknowledged.


\end{document}